# Integrable Magnetic Fluid Hyperthermia Systems for 3D Magnetic Particle Imaging

André Behrends[1#], Huimin Wei[1#], Alexander Neumann[2], Thomas Friedrich[1], Anna C. Bakenecker[1], Matthias Graeser[1,2] and Thorsten M. Buzug[1,2]

1. Fraunhofer Research Institution for Individualized and Cell-Based Medical Engineering (IMTE), Lübeck, Germany
2. Institute of Medical Engineering (IMT), University of Lübeck, Lübeck, Germany

# These authors contributed equally.
Corresponding authors: André Behrends, Email: andre.behrends@imte.fraunhofer.de; Huimin Wei, huimin.wei@imte.fraunhofer.de

## Abstract

**Background:**
Combining magnetic particle imaging (MPI) and magnetic fluid hyperthermia (MFH) offers the ability to perform localized hyperthermia and magnetic particle imaging-assisted thermometry of hyperthermia treatment. This allows precise regional selective heating inside the body without invasive interventions. In current MPI-MFH platforms, separate systems are used, which require object transfer from one system to another. Here, we present the design, development and evaluation process for integrable MFH platforms, which extends a commercial MPI scanner with the functionality of MFH.

**Methods:**
The biggest issue of integrating magnetic fluid hyperthermia platforms into a magnetic particle imaging system is the magnetic coupling of the devices, which induces high voltage in the imaging system, and is harming its components. In this paper we use a self-compensation approach derived from heuristic algorithms to protect the magnetic particle imaging scanner. The integrable platforms are evaluated regarding electrical and magnetic characteristics, cooling capability, field strength, the magnetic coupling to a replica of the magnetic particle imaging system's main solenoid and particle heating.

**Results:** The MFH platforms generate suitable magnetic fields for magnetic heating of particles and are compatible with a commercial magnetic particle imaging scanner. In combination with the imaging system, selective heating with a gradient field and steerable heating positioning using the MPI focus fields are possible.

**Conclusion:**
The proposed MFH platforms serve as a therapeutic tool to unlock MFH functionality of a commercial magnetic particle imaging scanner, enabling its use in future preclinical trials of MPI-guided, spatially selective magnetic hyperthermia therapy.

## Main Text

### Introduction and motivation

In the last years, magnetic modalities experienced an upswing in medical technology sciences. Especially, magnetic nanoparticles have shown large potential varying from imaging [1], drug delivery [2] and microrobotics [3] to hyperthermia [4], providing a large potential both in diagnostics as well as in therapy. Huge benefit is expected from combining diagnostic imaging with therapeutic applications. In such theranostic settings, the advantages of both sides can play out and enable new therapeutic approaches in complex clinical settings. An outstanding example for such a setting is a magnetic particle imaging (MPI) system with hyperthermia capabilities [4,5]. The combination of MPI with hyperthermia benefits in two ways. First, the selection field topology of an MPI device enables spatially confined energy deposition [6]. Second, the tissue temperature can be determined by multi contrast imaging ensuring successful treatment and protection of healthy tissue by preventing overheating [7]. The following part gives a brief overview of the two modalities MPI and magnetic fluid hyperthermia (MFH) as well as the combined usage of them. The presentation of the motivation and scope of this article closes this part.

**Brief introduction to magnetic particle imaging**

Magnetic particle imaging (MPI) is a tracer-based imaging modality first presented in 2005 by Gleich and Weizenecker [1]. MPI uses the nonlinear magnetization behaviour of magnetic nanoparticle tracers (MNPs) to map their spatial distribution. A common MNP tracer material is magnetite ($Fe_3O_4$). The tracer particles are driven to saturation by a homogeneous sinusoidal magnetic field (drive field) and, due to their nonlinear magnetization, respond with a signal containing higher harmonic frequency components. The time-dependent magnetization of the particles can be recorded using an inductive coil sensor. Spatial resolution is achieved by superposition of a gradient field (selection field) in addition to the drive field, which results in a spatial dependent field sequence and particle response. Solely, particles located in the vicinity of a field-free region in the gradients symmetry center, whereas this region may be a point (field-free point, FFP) or a line (field-free line, FFL) [8], provide a significant contribution to the signal. The spatially dependent particle responses are then reconstructed in a concentration image using a regularized minimization approach and a Kaczmarz algorithm. Moreover, the magnetization behaviour depends, among other parameters, on the temperature and the viscosity of the surrounding medium, which enables multi-contrast MPI to reconstruct temperature and viscosity maps [7,9,10].

**Brief introduction to magnetic fluid hyperthermia**

As far as we are aware, the first publication proposing the use of induction to locally heat tissue dates back as far as 1957, when Gilchrist et al. proposed the selective heating of lymph nodes for cancer therapy [11]. The working principle of magnetic fluid hyperthermia (MFH) is based on hysteresis losses in magnetic materials. If a magnetic material, like magnetic nanoparticles, is placed in an alternating magnetic field (AMF) the magnetic material will convert magnetic energy into thermal energy due to the hysteresis losses and consequently heat up the surrounding tissue [12,13]. Applications of MFH range from cancer treatment [14] to opening the blood brain barrier [15]. Magnetic fluid hyperthermia is not limited to preclinical application, but has already advanced into clinical scenarios [16].

**Combinations of magnetic particle imaging and magnetic fluid hyperthermia**

Since MPI and MFH both rely on the interaction of MNPs and AMFs the question whether both modalities can be combined arises. Moreover, the possibility of MPI to map temperatures inside the body is the perfect complementary evaluation modality for MFH. Furthermore, the principle used for spatial encoding in MPI can likewise be used to apply spatial selectivity in MFH. In contrast to MRI the field-free region in the symmetry center of the gradient field enables for full hysteresis loops of magnetic materials. As this region can also be steered with additional static fields, the region for localized MFH can be steered as well. For comparison, due to the strong $B_0$ field used in MRI particles would stay in saturation with achievable high frequency AMFs, thus making the combination of MFH and MRI unfeasible. Previous research already showed the feasibility of the combination of MPI and MFH [5,17]. However, a major drawback of current approaches is the application of MPI and MFH in separate devices. Thus far no single device or integrated system exists where MPI and MFH can be applied without moving the specimen. A system providing MPI and MFH capabilities without moving the specimen can either be constructed from scratch or as an extension to existing systems. This contribution aims for the latter approach of an integrated extension for a commercially available MPI scanner.

**Motivation and scope of the article: Design, construction, and preliminary technical testing of integrated magnetic fluid hyperthermia systems for magnetic particle imaging**

Until now, MFH devices do not work within the imaging volume of multidimensional MPI systems. To enable the full potential of the combined technologies we propose a hyperthermia enabled MPI device. An integrated operation may allow fast switching between imaging and heating modes and thus online evaluation of MFH, utilising multi-contrast MPI. As typical frequencies of MFH range from 100 kHz to 1 MHz, which is well within the MPI receive band, several difficulties must be addressed in the design process of integrated MFH-MPI systems. First, the receive chains must be protected against over voltage from the alternating magnetic fields by compensation or filtering. As signals in the receive chain are typically amplified by a factor of 40 dB to 60 dB in the receive band, a combination of compensation and filtering is advised to protect the sensitive low noise electronics. If operated fast enough, blanking circuits which disconnect the low noise amplifier (LNA) from the receive chain may be used as well. In total this paper describes the design for a combined operation of magnetic hyperthermia and MPI using integrable MFH systems. We present the design and development process, as well as evaluation of the integrable MFH systems.

## Material and methods

A solenoid coil is chosen as initial topology for the MFH system. The MFH system thus provides a magnetic field mainly along the bore direction of the MPI system. The basis for the development process of the integrable hyperthermia systems is the below discussed requirements specifications. The key to a successful integration is magnetic self-compensation, which is briefly presented before discussing the steps taken to develop self-compensating systems. The algorithms used to find suitable winding profiles for the systems are introduced and the self-compensation is evaluated in small-signal regime using rapid prototyping. Finally, the high-power integrable systems and the impedance matchings are presented.

**Design requirements for integrated hyperthermia systems in MPI**

A fundamental constraint is given by the bore of the magnetic particle imaging system. The bore diameter of the imaging system is 119 mm, thus the outer diameter of the MFH systems must be smaller. A more challenging issue is the cross-coupling of the systems, which can lead to induced voltages that may cause severe damage to either of the systems. Comparing MFH and MPI systems, the imaging system is more susceptible to damage caused by induced voltage, due to its use of sensitive electronics such as low-noise amplifiers. Therefore, the main goal of this work is to avoid destruction of the MPI system by the MFH system. One solution could be direct filtering of the signal in the receive chain of the imaging system. However, a filter with sufficient attenuation to lower the signal far enough to avoid damage will have strong impact on the receive sensitivity of the broad band receive chain. Therefore, filtering efforts must be kept as low as possible. An alternative is magnetic self-compensation to reduce the induced voltage in the MPI system by manipulating the magnetic flux densities of the systems properly. The critical quantity in the receive chain which leads to destruction of sensitive components is the voltage. The most critical component in the receive chain are the JFETs of the LNA. A typical JFET used in MPI is the BF862 [18], which has a breakdown voltage of 20 V. Therefore, we have chosen 16 V, including a safety margin of 20 %, as the maximum allowed induced voltage. Since access to the system is limited, direct measurement of the induced voltage is not possible. The LNA outputs are the first point within the receive chain which are accessible, thus all measurements were performed at the output of the LNA. Typical transfer-functions are measured as power ratios and the gain of the amplifier is given as power ratio as well. Because of this we chose to characterize the self-compensation by the allowable transferred power. The allowable transferred power is calculated from the allowed voltage by using the following relation

$$L_P = 10 \cdot \log_{10}\left(\frac{|S_{MPI}|}{|S_{MFH}|}\right) \geq 20 \cdot \log_{10}\left(\frac{|U_{MPI}|}{|U_{MFH}|}\right) = L_U. \qquad (1)$$

Where $S_{MPI}$ and $S_{MFH}$ are complex powers and $U_{MPI}$ and $U_{MFH}$ are the voltages of the MPI and MFH coils respectively. Since value of the power level $L_P$ is always bigger as the value of the voltage level $L_U$, setting the value of the voltage level as limit for the power level introduces a safe condition. For a derivation of (1) refer to the supplementary materials. The residual voltage may be further filtered using off-the-shelf components and comparatively low effort. Literature related to MFH suggests the use of frequencies in the range of 100 kHz to 1 MHz and field strengths of up to 12 mT for biological applications [14,19,20]. Since the self-compensation approach uses an additional field which cancels the magnetic flux, coupled in the receive chain, the induced voltage is reduced. This results in higher power consumption compared to MFH systems due to the additional turns. As a consequence of the high frequency and flux density requirements, no soldering joints are allowed in the integrable system, since the joints would experience excessive heating, caused by induced eddy currents. Thus, the coil must be wound from a single wire. Additionally, to allow for complementary imaging the setup must not contain any magnetic or large conducting materials, significantly limiting the choice of materials.

**Theory of magnetic self-compensation**

The chosen approach to integrate the MFH systems into the MPI system without causing damage is magnetic self-compensation. Magnetic self-compensation arises from the cancellation method [21,22] known from receive signal processing in MPI and uses Lorentz reciprocity to adapt this principle to magnetic field-generation. Self-compensation is achieved by compensation turns, which create a magnetic field cancelling the total flux in the receive coil, while the primary heating field still can perform MFH in the region of interest (ROI).

**Design approaches**

Initially, a simulation-based approach has been tested to develop the integrable MFH systems, which lead to a basic understanding of the problem and the following general rules: first, the heating turns should be placed as close as possible to ROI. Second, the compensation turns should be placed as far away from heating turns and ROI as possible, while still being able to induce voltage into the imaging system. Third, both heating turns and compensation turns should fit in the MPI system imaging bore and ensure enough free space for imaging objects.

Those rules ensure reduction of power demand while having effective compensation and yield the basic coil topology shown in figure 1. The heating turns form a part of the MFH systems which will be referred to as heating winding, whereas the compensation turns form a part of the MFH systems referred to as compensation winding. The further design process uses a measurement-based approach.

**Measurement-based approach**

To gather data which includes all influences of the target system a measurement-based approach following [22] has been used. For this approach a single turn coil probe is moved step-wise along the scanner bore (x direction) and the voltage induced by the scanner into the coil probe is measured with an oscilloscope at each position. The measurement process is depicted in figure 2. The procedure is conducted separately using coil probes with the respective diameters of the heating and compensation turns. The measurements were performed during three 1-D MPI imaging sequences, one for each channel and its corresponding spatial directions (x, y and z). The procedure yields the field profiles $\hat{u}_x^h$, $\hat{u}_y^h$, $\hat{u}_z^h$ $\hat{u}_x^c$, $\hat{u}_y^c$ and $\hat{u}_z^c$ with respect to the coil probes position in the scanner, where the superscript indicates heating (h) or compensation (c) winding and the subscript indicates the channel x, y or z. Using the law of reciprocity one can deduct a winding profile which minimizes the induced voltage from the MFH system to the MPI scanner. The field profiles have been recorded using N = 201 spatial positions. The field profiles for the x-, y- and z-channel are shown in figure 3. The asymmetry of the field profiles, especially in the y- and z-channel, may be caused by magnetic or conductive parts as well as connection wires in the MPI system. The asymmetry can become apparent because the heating and compensation turns are much closer to the bore walls and at a considerable distance to the MPI imaging volume.

## Optimization of winding profiles

Initially, an algorithm based on preconditioned iterative reduction by absolute value (PI-RAV) has been implemented and used. Even though this algorithm finds suitable solutions the algorithm is not robust. Therefore, a modified version of the differential evolution particle swarm optimization algorithm (mDEPSO) is developed to further optimize the generation of the compensation winding [23].

**Statement of minimization problem**

The task at hand is to place heating turns and compensation turns in a way that the induced voltage to the MPI system is minimal or at least on an acceptable level. The collection of directions and positions of the wire turns for each winding is called a winding profile. Let $w_h$ and $w_c$ be the winding profiles of the heating and compensation winding respectively. Given the measured field profiles $\hat{u}_x^h$, $\hat{u}_y^h$, $\hat{u}_z^h$, $\hat{u}_x^c$, $\hat{u}_y^c$, $\hat{u}_z^c \in \mathbb{R}^N$, N = 201. The problem can be stated as:

$$\min_{\vec{w^j}} \sum_i \left| \sum_j w^j \cdot \hat{u}_i^j \right| ; \qquad i \in \{x, y, z\}; \qquad j \in \{h, c\} \qquad (2)$$

It is useful to reduce the degrees of freedom for such a problem using rational arguments. It makes sense to allow only a single winding direction for the heating winding, to avoid self-compensation of the heating field itself, but allow both directions for the compensation winding, to increase the freedom of choice for compensation. This can be modeled as $w_k^h \in \{0, 1\}$ and $w_k^c \in \{-1, 0, 1\}$ with k = 1, ..., N. Nonetheless, the problem is of high dimension and its solution space is not guaranteed to have any global minimum, since this is highly dependent on the recorded field profile.

**Preconditioned iterative reduction by absolute value**

The preconditioned iterative reduction by absolute value algorithm (PIRAV) is based on an iterative compensation for the heating winding. For a fixed number of heating turns and a predefined allowable area for compensation turns, in conjunction with a fixed cable width, the algorithm chooses the compensation turn position and direction which minimizes the absolute value of the voltage in each iteration and removes the chosen position from the set of possible compensation turn positions. This process repeats until no further improvement is achieved or a maximum number of compensation turns is reached. Additionally, this process repeats for an increasing number of heating turns, starting with a single turn as close as possible to the ROI up to an arbitrarily chosen number of turns. From all the solutions for the different numbers of heating coil turns the one with the lowest power consumption is used.

**Modified differential evolution particle swarm optimization**

Preliminary to this work, we developed an optimization algorithm with differential evolution (DE) operator based on simple particle swarm optimzation (sPSO). sPSO has an improved convergence efficiency, compared to classical PSO, but is still easily trapped in local extrema [24]. The DE operator introduces random mutations to the particle swarm to increase the population variety [25], which gives the proposed mDEPSO method a better performance. In mDEPSO, the particles are first manipulated by the DE operator, if there is no fitness improvement, the sPSO will be performed on the particles [26].

**Small-signal evaluation of self-compensation**

A small-signal prototype of the integrable MFH system was built, consisting of the windings placed on a minimal framework, according to the calculated winding profiles. The framework of the prototype has been 3D printed on an UltiMaker 2 (Ultimaker B.V., Utrecht, Netherlands) and a Form 2 (Formlabs Inc., Somerville, United States of America) printer. The purpose of the prototypes is the evaluation of the transfer-functions from the MFH systems to the MPI system, as well as the estimation of their electrical and magnetic characteristics. From the electrical characteristics the voltage across the coil U of the MFH systems can be estimated from the complex power of the coil using the amplifier power P as

$$|U| = \sqrt{S \cdot |Z|} = \sqrt{P \cdot |Z|},$$

where Z is the impedance of the MFH systems' coil. Using equation 1, the power ratio is given by

$$L_P = 10 \cdot \log_{10}\left(\frac{|S_{MPI}|}{|S_{MFH}|}\right) = 20 \cdot \log_{10}\left(\frac{16\,V}{\sqrt{S \cdot |Z|}}\right). \tag{6}$$

**Transfer-functions**

To evaluate the principle of magnetic self-compensation, the power transfer from the output of the power amplifier to the coils of the MPI scanner through the MFH system is investigated. It is measured over a frequency range from 100 kHz to 1 MHz and compared to the MFH system without compensation winding. The measurement is carried out using a network analyzer (Keysight E5061B; Keysight Technologies Inc., Santa Rosa, United States of America), using T/R-measurement mode. The test signal of the network analyzer is fed to the small-signal prototype, couples inductively into the receive coils and is measured at the output of the LNA within the MPI receive chain. The measured transfer-functions must be corrected for the gain of the low-noise amplifier of the MPI system, which is around 40 dB according to the manufacturer and an additional matching factor of

$$\frac{P_{\text{MPI}}}{P_{\text{Test}}} = \frac{\sqrt{R_L^2 + (\omega L)^2}}{R_L}. \tag{7}$$

The correction factor given in equation 7 needs to be applied since the small-signal evaluation, giving $P_{\text{Test}}$, is performed using an unmatched coil while actual MFH, giving $P_{\text{MPI}}$, will be performed with a matched coil. A derivation of equation 7 can be found in the supplementary materials. The transfer-function of an MFH system small-signal prototype, corrected for low-noise amplifier gain and matching factor, is shown in figure 4. For the shown case the allowed voltage level equates to -16.57 dB, according to equation 6. The transfer-functions of the heating winding and the self-compensated MFH system show a decrease of overall transferred power. As mentioned before, it can be seen from figure 4, the main effect of self-compensation occurs in the x-channel (along the MPI scanner bore). Even though the y- and z-channel show a slight increase in transferred power, the power levels are low and especially for the frequency chosen for MFH marked in green the transferred power levels are on a sufficiently low level, since the gain/loss level is below the target of -16.57 dB for all channels.

**Integrable prototypes**

### Construction

The final design of the integrable MFH system adopts the winding profiles of the small-signal prototype, since the result of the small-signal transfer functions were positive. The integrable MFH system embodies a cooling system for the windings to enable application of long duty cycles at high currents gaining as much field strength as possible. Transparent material is used for the housing to ensure the cooling system is free of air bubbles by visual inspection. The whole MFH system including the cooling is designed to fit into the MPI systems bore.

### Evaluation of magnetic self-compensation

To evaluate the self-compensation capability of the MFH systems, the main solenoid of the commercial MPI system (preclinical MPI 25/20 FF Scanner, Bruker BioSpin MRI GmbH) has been replicated. The MFH systems are operated with a power of 10 W and moved along the axis of the solenoid coil replica in steps of 2.5 mm, thus inducing a voltage in the replicated solenoid depending on the position of the MFH system. This yields an induction profile, dependent on the offset of the ROI of the MFH system and the geometric center of the solenoid replica.

### Evaluation of cooling capability

To verify safe operation limits with respect to field generator cooling, different powers have been applied to the MFH systems while measuring the temperature of the coils. The conditions to end the cooling test are a stable temperature over a time of 3 minutes or the coil reaching a defined maximum temperature. In case the temperature limit is reached, the time until the temperature has been reached is noted. The recorded temperature data for the evaluation of the cooling capability can be found in the supplementary materials.

### Evaluation of electrical characteristics and magnetic field strength

To further analyze the final MFH systems, their electrical characteristics, namely the induction and the series resistance must be measured. This is done using a precision LCR meter (Keysight E4980A).

To specify a precise value of the achievable magnetic field strength an experimental evaluation of the MFH systems is necessary. A DC source supplies the current $I_{max}$ flowing through the field generator at the cooling-limited maximum power $P_{RMS}$, a hall probe is used to measure the maximum magnetic field amplitude $B_{max}$. The current depends on the maximum power that can be fed into the field generator as well as the resistance of the coil. The current amplitudes can be calculated using the resistance $R_S$ of the field generator coil

$$I_{max} = \sqrt{\frac{2 \cdot P_{RMS}}{R_S}}.$$

### Evaluation of particle heating

To evaluate the capability of the MFH systems to heat magnetic nanoparticles, particles are placed inside the MFH field generator and a magnetic field is applied. 200 μL undiluted Resovist© (Bayer Pharma AG, Berlin, Germany) with an iron concentration of 28 mg mL$^{-1}$ has been used to evaluate the particle heating. All measurements are performed at the respective cooling-limited maximum power of the MFH system.

**Impedance matching**

To ensure maximum power transfer from the power amplifier to the field generator of the MFH system, the system is matched to the power amplifiers optimal load by impedance matchings.

**Design and construction**

The impedance matching is a type LCC impedance matching consisting of an inductor L and two capacitors C as depicted in figure 5. The inductor of the LCC impedance matching is the field generating coil, which is matched to the power amplifiers optimal load by a series capacitor and a parallel capacitor. This type of impedance matching ensures optimal energy transfer from power amplifier to the field generator for a single frequency. The frequency is chosen as the heating frequency of 700 kHz. The reactance of the series capacitor $X_{CS}$ and the parallel capacitor $X_{CP}$ can be calculated using the following formulas

$$X_{CS} = \sqrt{R_S}\sqrt{Z_{src} - R_S} - X_{LS}$$

and

$$X_{CP} = -\sqrt{R_S}\frac{Z_{src}}{\sqrt{Z_{src} - R_S}},$$

where $R_S$ and $X_{LS}$ are the resistance and the reactance of the magnetic field generating coil, respectively. The optimal load impedance of the power amplifier is given by $Z_{src}$. The capacitance values for the series and parallel capacitor can be achieved from their respective reactances. The capacitors are high power conduction cooled capacitors (Celem Passive Components Ltd., Jerusalem, Israel) directly installed on copper plates for minimal contact resistance. The copper plates are cooled with oil and serve as heat sink. The impedance matching is cooled in series with the coil and is cooled first due to higher temperature rise of the coil. Low resistance connections are intended to limit resistive losses in the system, whereas low inductance connections avoid degrading of the impedance matching while operating at high powers.

# Results and discussion

Two MFH systems have been built according to the procedures presented in the materials and methods section. One system is capable to fit a rat and uses a single-coil heating winding, the other system is capable to fit a mouse and uses a split-coil heating winding.

**Winding profiles**

The winding profiles for the systems are shown in figure 6. The mDEPSO algorithm was used for the single-coil MFH system and the PIRAV algorithm was used for the split-coil MFH system to generate the winding profile. Due to the coil splitting, the split-coil system can be equipped with an auxiliary device, e.g. a high intensity focused ultrasound (HIFU) transducer, the installation and combination of the splitted MFH system is beyond the scope of the here presented work and will be part of future developments. As might be expected, the asymmetries observable in the field profiles are emerging in the winding profiles as well.

**Cooling**

Working under stable conditions is possible for the single-coil MFH system for up to 250 W and for the split-coil MFH system for up to 150 W. Under unstable conditions the MFH systems might be operated in duty cycles. For powers leading to unstable cooling conditions the heating time is defined, as the time until the coil reaches its temperature limit, and the cooling time, as the time the coil needs to cool back to ambient temperature. The maximum powers to operate in duty cycles, as well as the heating time and temperature limit of the systems can be found in table 1. The final single-coil MFH system with integrated cooling is shown in figure 7, the split-coil MFH system in figure 8.

**Magnetic field strength**

For the single-coil MFH system the cooling-limited maximum power is 600 W, thus the magnetic flux density in the ROI, for operation in duty cycles is 10 mT. For continuous operation the maximum power is 250 W, resulting in a magnetic field strength of 6.4 mT.

According to the cooling capability test the maximum power for the split-coil MFH system is 350 W. The resulting achievable magnetic field strength in the ROI of the split-coil MFH system is therefore 4.18 mT for operation in duty cycles. For continuous operation the power is 150 W, corresponding to a magnetic flux density of 2.7 mT.

**Magnetic self-compensation**

The induced voltage profiles, showing the voltage induced in the replica of the imaging systems main solenoid, depending on the position of the MFH systems, are shown in figure 9. The profiles have been normalized by the magnetic flux density of the MFH systems in their ROI. Intuitively, one would expect the minimum to be at the center. However, this is only true if the FFP position of the MPI system is equal to the geometric center of the solenoid coil replica, which cannot be assured. Additionally, manufacturing variations or eddy currents within the imaging system may be a source of error which are not reproduced in the replica coil, altering the profile for the induced voltage.

The upper graph of figure 9 shows, that the minimum induced voltage per magnetic flux density for single-coil MFH system is at an offset of around 14 mm with an induced voltage per magnetic flux density of 13 V mT$^{-1}$. This can be explained by the larger size of the cancellation turns which interact stronger with shielding faces around the bore.

The minimum induced voltage per magnetic flux density is at an offset of around 7 mm from the center for the split-coil MFH system, as can be seen from the lower graph of figure 9. The exact minimum has been determined by manually shifting the system around this offset and is found to be at 6.5 mm with an induced voltage per magnetic flux density of 1.019 V mT$^{-1}$.

Both MFH systems are capable of self-compensation as it is evident from figure 9. However, the self-compensation of the split-coil MFH system is more efficient showing a minimum in the induction profile of 1.019 V mT$^{-1}$ compared to 13 V mT$^{-1}$ for the single-coil MFH system. For the split-coil MFH system with a magnetic flux density of 4.18 mT the induced voltage in the coil replica is 4.26 V, well below the allowed voltage of 16 V. For a magnetic flux density of 10 mT for the single-coil MFH system, this equals an induced voltage in the coil replica of 130 V. Therefore, a separate filter is needed with a loss level of

$$20 \cdot \log_{10}\left(\frac{16 \text{ V}}{130 \text{ V}}\right) = -18.2 \text{ dB}.$$

Which can be easily implemented as a second-order band-stop filter. However, the sensitivity of the self-compensation with respect to the position of the MFH inserts is much lower for the single-coil MFH system, since the steepness of the induction profile is much lower compared to the induction profile of the split-coil MFH system.

**Particle heating**

Measurements using an infrared thermal camera was performed for the split-coil MFH system and served as a first indicator of proper particle heating (see figure 10). Due to the closed bore geometry of the single coil MFH system, these measurements were not possible for this coil geometry. Figure 11 shows the temperature curves of 5 heating cycles for both MFH systems in duty cycle mode, measured with a fiber optical thermometer.

For the single-coil MFH system, the temperature increase during a single heating cycle reaches up to 21.3 K and over 5 heating cycles a temperature rise of 28.5 K can be achieved (figure 11, upper graph).

For the split-coil MFH system, the temperature increase during a single heating cycle reaches up to 7 K and over 5 heating cycles a temperature rise of 13.4 K can be achieved (figure 11, lower graph).

Both MFH systems can heat the particles significantly. In comparison, the single-coil MFH system can increase the temperature by 21.3 K per heating cycle, whereas the split-coil MFH system can increase the temperature by 7 K per heating cycle, which is a factor of 3 between the systems. This result is plausible, since the single-coil MFH system can produce more than twice the field strength compared to the split-coil MFH system.

**Specific absorption rate**

The specific absorption rate (SAR) of the particles can be obtained from the measured particle temperature change rate using:

$$SAR = \frac{C_w m_w + C_p m_P}{m_p} \cdot \frac{\Delta T}{\Delta t}$$

where T is the measured temperature, t is the elapsed time, $C_w = 4190$ J kg$^{-1}$ K$^{-1}$ is the specific heat capacity of water, $C_p = 670$ J kg$^{-1}$ K$^{-1}$ is the specific heat capacity of the particle [5], $m_w$ and $m_p$ are the masses of water and magnetic particles in the sample, respectively. The temperature change rate $\Delta T/\Delta t$ should be measured under adiabatic conditions, which is experimentally unachievable because minimal heat transfer is inevitable across the sample-environment boundary. An approximation method [27] is used to identify the quasi-adiabatic heating conditions in the heating experiment. The SAR value for the for the single-coil MFH is 60 Wg$^{-1}$ and the SAR value split-coil MFH system is 10 W g$^{-1}$. The derived SAR values confirm a 6-fold increased MFH capability for the single-coil MFH system compared to the split-coil MFH system.

**Conclusion and outlook**

### Brief summary and system comparison

The design process of two MFH systems which can be integrated in a commercial MPI system has been presented. The integrable prototypes have been evaluated for their electrical and magnetic characteristics, as well as their capability to heat particles and their magnetic self-compensation. The split-coil MFH system provides additional space for auxiliary devices such as a HIFU transducer, giving direct access to the ROI and shows good self-compensation capability inducing only around 1V mT$^{-1}$ into the imaging system. A drawback of the split-coil MFH system is its limited capability in generating magnetic field strength, being able to produce a magnetic flux density of around 4.2 mT in the ROI and thus limited heating capabilities indicated by the SAR of 10 W g$^{-1}$ for the used particles. The single-coil MFH system provides limited access to the ROI and less self-compensation capability with an induced voltage of around 13 V mT$^{-1}$ into the imaging system. However, the single-coil MFH system can produce a higher magnetic flux density of around 10 mT in the ROI and thus provides higher heating efficiency indicated by the SAR of 60 W g$^{-1}$, using the same particles as used for the estimation of the SAR for the split-coil MFH system.

The results in this work indicate a trade-off between power demand and self-compensation exists which needs to be validated in further studies.

Both systems show clear strengths, but also opportunities for improvement, which will be discussed in the following sections.

**Preceding works, fundamental requirements and their implications for the presented systems**

*Comparison of preceding MFH-MPI approaches*

Few preceding works on combining MPI and MFH have been published, even though the relevance for their combination is acknowledged [6].

The first and single practical implementation of an MPI-MFH combination, known to the authors, relies on separate devices to combine multi-dimensional MPI and MFH, making it necessary to move the object of investigation between the two modalities [5]. Additionally, the existing system operates at the same frequency for MPI and MFH of around 300 kHz, however MPI is recommended only up to 150 kHz [28].

Wells et al. were able to perform MFH with the native MPI frequency of 25 kHz. Using particles with an iron concentration of 935 mmol l$^{-1}$ and magnetic field strengths of 12 mT they were able to achieve SAR values of up to 2.5 W g$^{-1}$ [29]. However, further improvement of the heating efficiency is limited by the imaging system's technical capabilities.

In a different approach He et al. presented a simulation study for a combined MPI-MFH device engaging the task by building a new device from scratch, which is supposed to be capable of imaging and hyperthermia without the need of integrating an MFH system into the imaging device [30]. This approach would be an easy-to-use system, since placement and alignment of the MFH systems becomes unnecessary. However, an approach using integrable systems, as presented in this paper, allows for more flexibility, since the different systems can be exchanged depending on the specific needs.

Using the approach presented in this paper imaging can be performed using the MPI systems native frequencies of around 25 kHz and hyperthermia can be applied using a more convenient frequency for MFH, which is 700 kHz in this work. Due to the integrability of the presented MFH systems into an imaging system, the need to move the object of investigation between the modalities is obsolete, rendering them first-of-its-kind systems for on-the-spot switching between MPI and MFH. Lately, the integration of the single-coil MFH system in a commercial MPI system has been shown [31] and further supports the claims of this paper. Since this paper focus on the development of integrable MFH systems, the reader is directed to [31] showing experiments on MPI-based thermometry and spatially selective MFH. The integrable MFH systems presented in this paper are the foundation of the forementioned publications, enabling the integration of MFH in a preclinical MPI system. This paves the way for first preclinical trials of MPI-guided, spatially selective MFH therapy.

*Limits of the magnetic field strength*

Magnetic field strength is one of the parameters for MFH controllable by technical systems. As is evident from this work: for similar particles and frequency the single-coil MFH system with a magnetic field strength of 10 mT can deploy more energy in terms of heat compared to the 4.18 mT of the split-coil MFH system, reflected in the respective SARs of 10 W g$^{-1}$ and 60 W g$^{-1}$.

This immediately raises the question for a possible increase of field strength for heat deposition. However, increase of magnetic field strength leads to a saturation in the SAR, as observed by Kerroum et al. [32] or Iacovita et al. [33]. Therefore, the gain due to an increase of magnetic field strength is limited.

Another limiting factor for the magnetic field strength in medical MFH application arises from patient safety. The magnetic fields must not harm the patient. Thus, a discussion on safety limits for medical MFH application is necessary. According to the "Guidelines for limiting exposure to electromagnetic fields (100 kHz to 300 GHz)" of the International commission on non-ionizing radiation protection [34] as well as the "1999/519/EC: Council Recommendation of 12 July 1999 on the limitation of exposure of the general public to electromagnetic fields (0 Hz to 300 GHz)" of the Council of the European Union [35], only vague information are given. The documents distinguish between whole body SAR and local SAR. For medical MFH localized exposure SAR is the most applicable value for a therapeutic scenario. For local exposure averaged over more than 6 minutes, both documents recommend a limit of 2 W kg$^{-1}$ for head and torso and 4 W kg$^{-1}$ for limbs in general public. In [34] at least recommendations for occupational exposure are mentioned. For the occupational scenario the recommended SAR limits are 10 W kg$^{-1}$ for head and torso as well as 20 W kg$^{-1}$ for limbs. However, these SAR values are highly dependent on the tissue and the limits are not intended for therapeutic scenarios. Hence, an informational and regulatory gap for therapeutical applications in the frequency range between 100 kHz and 1 MHz exists and must be addressed.

For in-vivo application preliminary works refer to the Atkinson–Brezovich limit [36], a product of frequency and magnetic field originally given as $4.85 \times 10^8$ Am$^{-1}$s$^{-1}$. Recent publications tend to extend this product up to a value of $9.46 \times 10^9$ Am$^{-1}$s$^{-1}$ [37], which for a frequency of 700 kHz would allow a magnetic field strength of 17 mT.

*Comparison of SAR in other MFH systems*

To assess the SAR values measured with the integrable MFH systems a comparison to SAR values of previous hyperthermia systems will be given. Note, that SAR depends on field strength, frequency of magnetic field and a multitude of properties of the used ferrofluid. The parameters influencing the SAR value greatly differ between the presented and discussed systems. However, the SAR is the common measure of energy absorption per unit mass during an application, giving insights in the heating capability of the presented system.

Gneveckow et al. characterize their clinical MFH system MFH300F reporting SAR$_{fe}$ values up to 35 Wg$^{-1}$ [38] using a dispersion of magnetite and maghemite nanoparticles with an iron concentration of 117 g l$^{-1}$ at a frequency of 100 kHz and a field strength of around 21 mT. They argue a minimum sample volume of 5 ml is necessary to evaluate the SAR value, since heat transfer to the surrounding is of greater importance for smaller samples. Zhang et al. report values of 4.5 W g$^{-1}$ for uncoated magnetite nanoparticles and 75 W g$^{-1}$ for coated magnetite nanoparticles at a frequency of 55 kHz and a magnetic field strength of 20 mT [39]. Phadatare et al. find up to 25 W g$^{-1}$ at a frequency of 265 kHz and a field strength of 3.77 mT for their SAR-optimized material $CoFe_2O_4@Ni_{0.5}Zn_{0.5}Fe_2O_4$, exploiting coupling of the magnetic hard material $CoFe_2O_4$ to the magnetic soft material $Ni_{0.5}Zn_{0.5}Fe_2O_4$ [40].

The given examples confirm the ability of the presented MFH systems to produce SAR values typical for MFH applications. Additionally, the impact of the particle properties on MFH performance becomes evident, especially from the work of Phadatare et al.. Considering the presented heating tests have been performed using widely used ferrofluid, not particularly suitable for MFH, huge potential for an increase in SAR lies in the use of carefully designed ferrofluids.

**Outlook**

The previous considerations revealed the potential for further improvement of the integrable MFH systems. A more powerful cooling would allow higher magnetic field strengths enabling the optimal selection of magnetic field frequency and magnetic field strength for maximum SAR. However, the useful increase in field strength is limited, as discussed in "Limits of the magnetic field strength".

Additionally, the seeming trade-off between capability of magnetic self-compensation and power demand should be addressed. Finally, the systems should be improved by increasing their usability. Providing tools and guides for installation can avoid misplacement of the MFH systems in the MPI system and ease the effort of installation, making the system better approachable to preclinical researchers.

Apart from the systems, the careful design of particles bears great potential for improving the heating capability in MFH which has become evident in the considerations in "Comparison of SAR in other MFH systems". Blending MPI-suitable and MFH-suitable ferrofluids might be promising for multimodal theranostic, since particle blends have been shown to be functional for MPI [41].

A major difficulty which needs to be addressed in future is the upscaling to human applications similar to MPI [42], since power demand increases severely for larger therapeutic volumes. An intermediate step towards full size human application of combined MFH and MPI might be the application to limited volumes like limbs or head, which already bear great clinical potential (e.g. treatment of brain tumors [43], bone metastases [44], osteosarcoma [45] or reversible opening of the blood-brain barrier (BBB) for drug delivery [46] and theranostics for neurodegenerative diseases [47]).

The presented MFH systems are set up to be used for studies on medical applications and biological specimens. We propose two preclinical applications that could be conducted with the MFH systems as presented in the paper, to encourage translation of combined MPI and MFH into preclinical usage.

*Reversible opening of the blood brain barrier*

The BBB is a unique property of the microvasculature of the central nervous system. It tightly regulates the movement of ions, molecules, and cells between the blood and the brain. The permeability of the BBB can be increased by magnetic heating and recovers after removing the magnetic field [15]. The single-coil MFH system is able to fit an average sized rat brain to locally increase the temperature of the brain tissue [23], which makes it suitable for reversibly opening the BBB with high spatial precision.

*Tumor ablation with HIFU support*

The split-coil MFH system can be equipped with a high-intensity focused ultrasound (HIFU) device, which can be used to preheat the tumor tissue, aiming at tumor destruction. Due to HIFU preheating much less additional temperature rise is necessary, enabling a better heating effect even at lower magnetic field strengths. For this scenario, an MPI-compatible HIFU transducer must be placed as an auxiliary device into the integrable MFH system [48].

# Abbreviations

AMF alternating magnetic field
BBB blood brain barrier
DE differential evolution
FFL field free line
FFP field free point
HIFU high-intensity focused ultrasound
LPF low pass filter
MFH magnetic fluid hyperthermia
MNP magnetic nanoparticle
MPI magnetic particle imaging
SAR specific absorption rate
PIRAV preconditioned iterative reduction by absolute value
PSO particle swarm optimization
ROI region of interest
RTD resistance temperature detector

## Competing Interests

The authors have declared that no competing interest exists.

# Tables

*Table 1: Cooling capability of the MFH systems*

|  | **single-coil MFH system** | **split-coil MFH system** |
|---|---:|---:|
| stable power | 250 W | 150 W |
| interleaved power | 600 W | 350 W |
| interleaved heating time | 90 s | 120 s |
| temperature limit | 50 °C | 60 °C |

# Figures

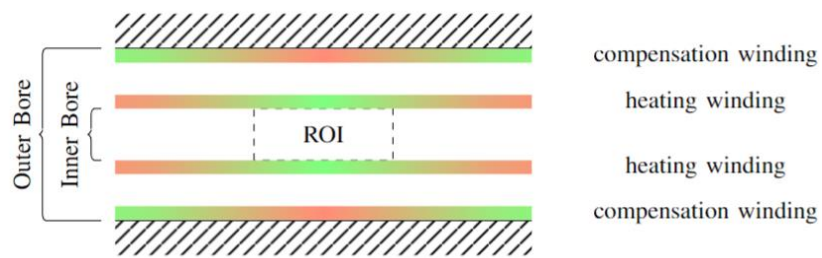

*Figure 1: Visualization of general placement rules: green areas indicate places where coil turns should be placed, red areas indicate places where coil turns should be avoided.*

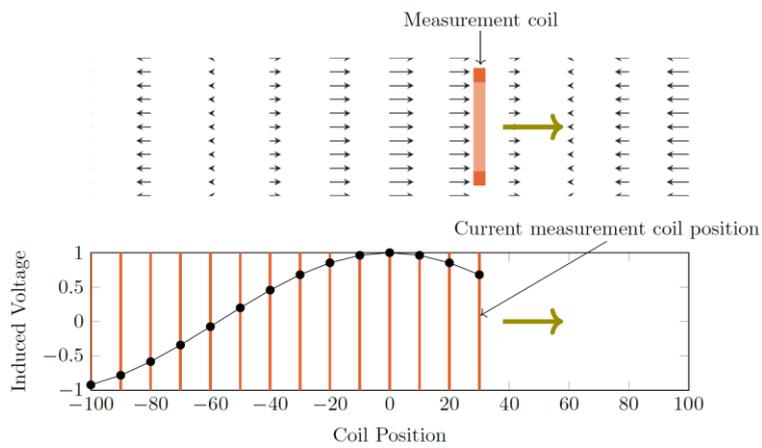

*Figure 2: A visualization of the induced field profile measurement. The upper image depicts the measurement coil (orange) moving through a non-homogeneous magnetic field (black arrows) in the direction of the green arrow. The lower image shows the amplitude of the induced voltage depending on the coil position, shown as orange lines. The images show a snapshot within the measurement process.*

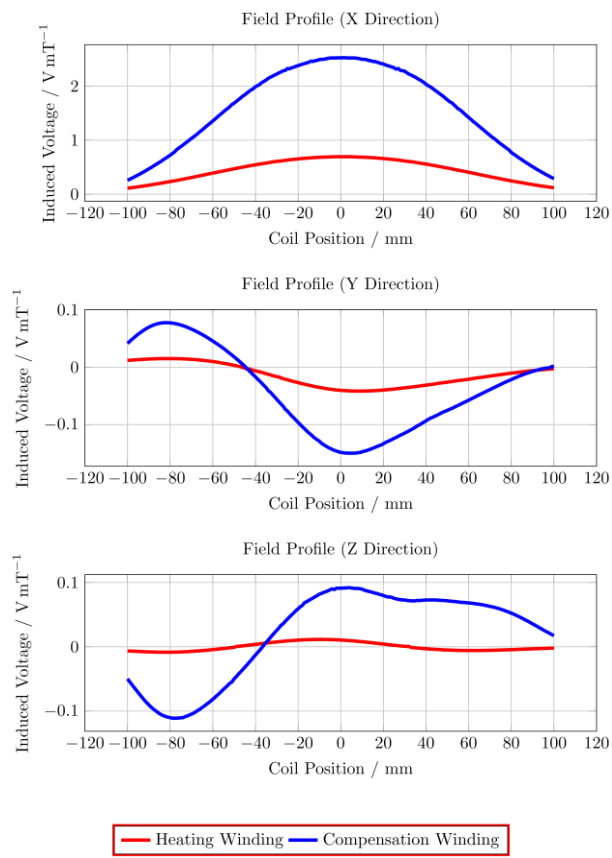

*Figure 3: The ratio of the induced voltage to the magnetic flux density is shown for a single compensation turn (blue) and a single heating turn (red) translated along the bore (x direction), while the MPI system creates a magnetic field along a certain direction (x, y or z), using the winding diameters for the MFH system.*

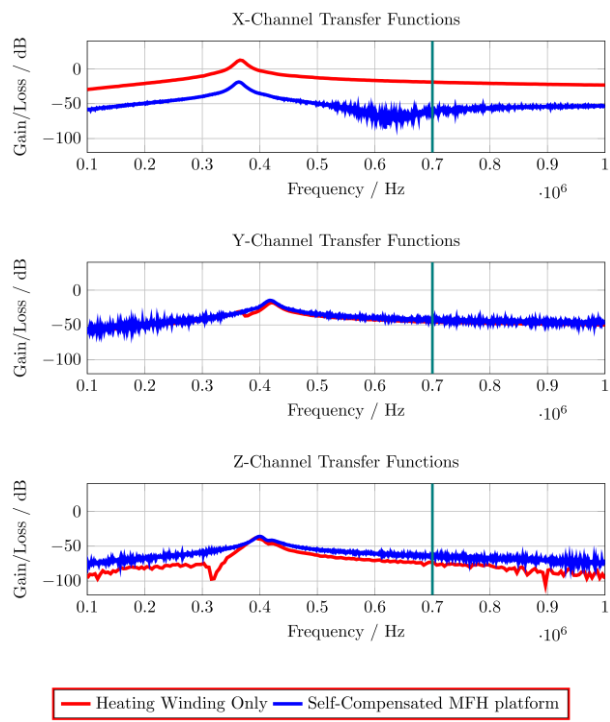

*Figure 4: The transfer-functions of the heating winding and the self-compensated MFH system show a decrease of overall transferred power. At the frequency chosen for MFH (marked in green), the transferred power levels areis below the target of −16.57 dB for all channels.*

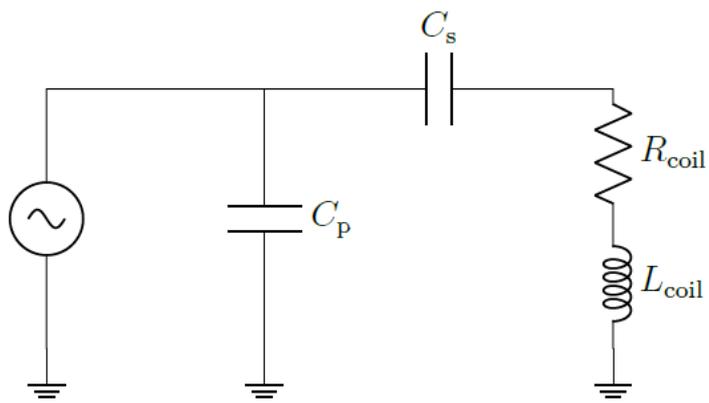

*Figure 4: Topology of the LCC impedance matching network, consisting of the inductor L with a resistance R, a series capacitor $C_S$ and a parallel capacitor $C_P$.*

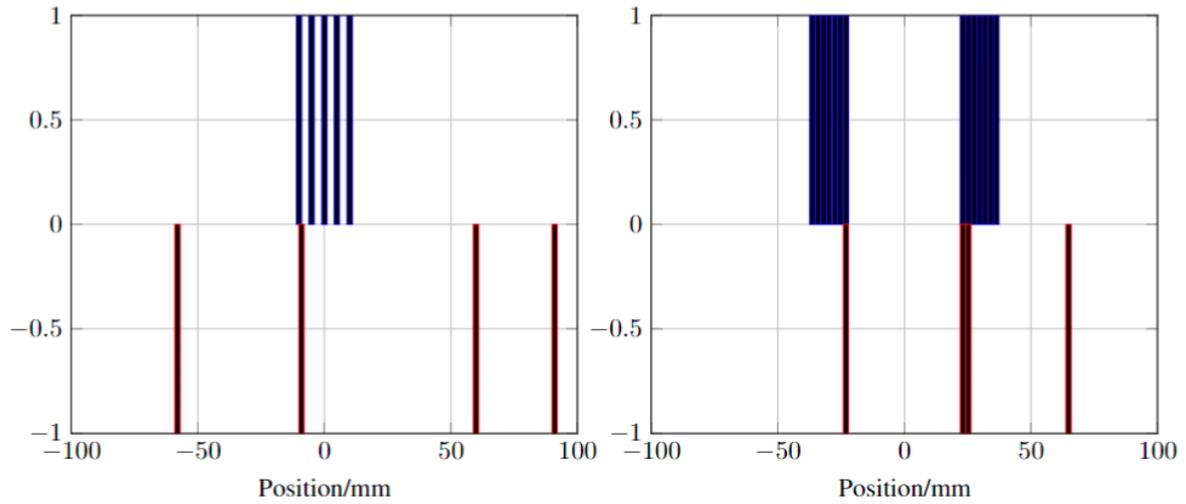

Figure 6: The winding profiles for the single-coil MFH system given by the mDEPSO algorithm (left) and the winding profile for the split-coil MFH system given by the PIRAV algorithm (right). The heating winding is plotted in blue and the compensation winding is plotted in red. The values of the winding indicate the direction of the coil turn, 0 means no turn is placed, -1 and 1 are opposing turns.

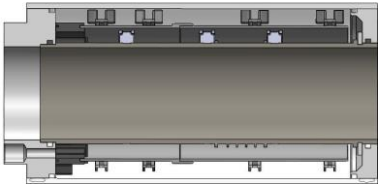

*(a) Section View of the single-coil MFH system CAD model*

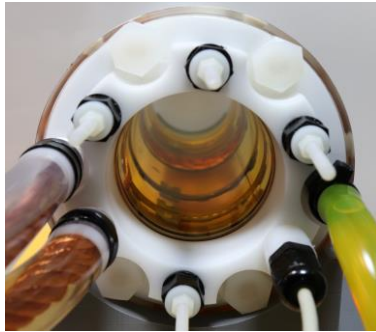

*(b) Front View of the single-coil MFH system.*

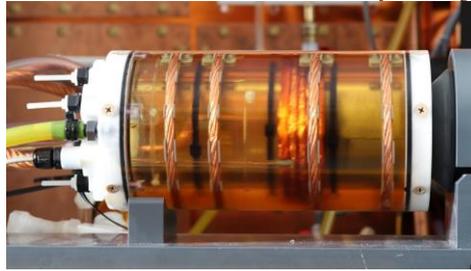

*(c) Single-coil MFH system placed on coil support cooled with oil.*

*Figure 7: Final design of the single-coil MFH system. The design of the cooling unit is shown in (a), yellow arrows indicate the direction of the cooling oil. (b) and (c) show the implemented single-coil MFH system. The coil winding as well as the temperature sensor can be seen through the transparent outer tube.*

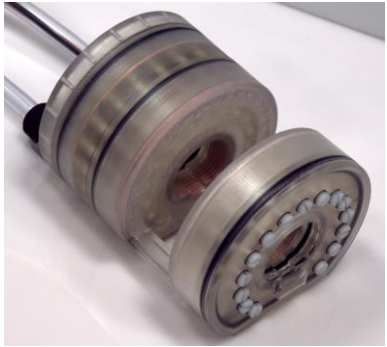

*(a) Top View of the split-coil MFH system.*

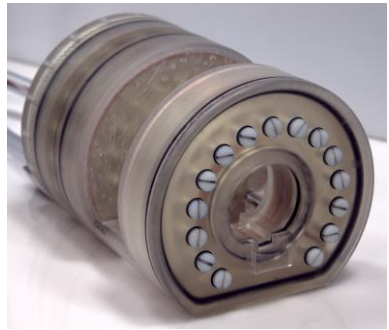

*(b) Front View of the split-coil MFH system*

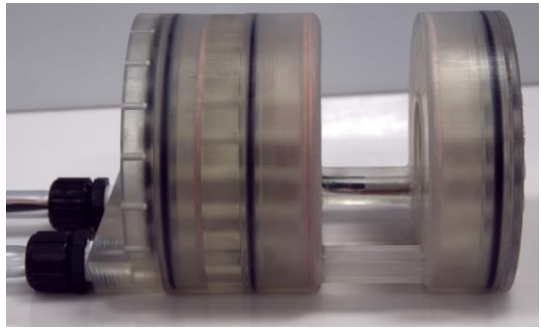

*(c) Side View of the split-coil MFH system*

*Figure 8: In (a) the overall split-coil MFH system can be seen, including the tubes for cooling and the electrical wires running within the cooling tubes. In (b) the inner and outer bore can be seen, as well as the flattened bottom, which is necessary for the wire placement of the auxiliary devices. In (c) the gap for the auxilliary devices is visible. Additionally, the coil wires (copper) and the o-rings (black) which establish tightness of the split-coil MFH system are perceivable.*

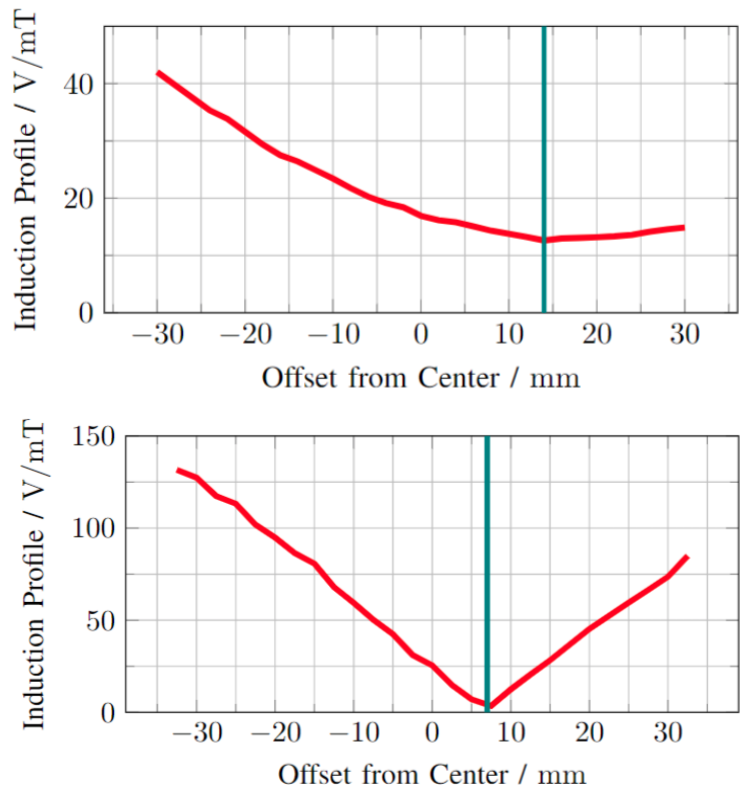

*Figure 9: The induced voltage, normalized by the center magnetic flux density of the MFH systems, in a replica of the scanners x-direction coil is measured depending on the offset between the geometric center of the scanner coil and the ROI of single-coil MFH system (upper) and the split-coil MFH system (lower) to each other along the bore direction.*

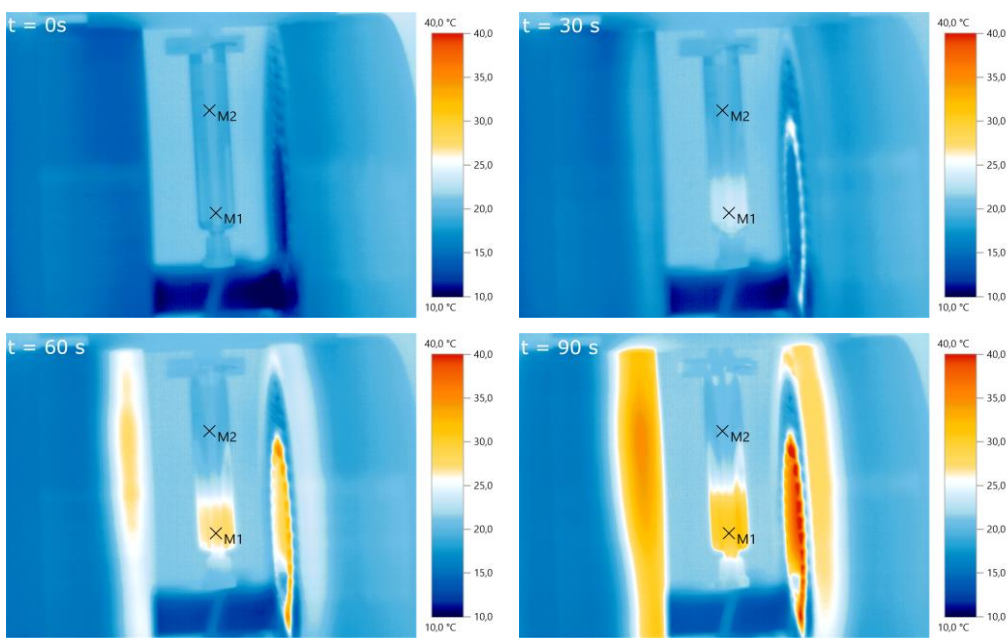

*Figure 10: Pre-evaluation of particle heating using infrared imaging for different times. A particle sample is placed in the center of the slit-coil MFH system and a magnetic field of xx mT is applied. The particles heat up from about 20°C to 30°C within 90 s. $M_1$ and $M_2$ are the measurement points of air and the particle solution, respectively.*

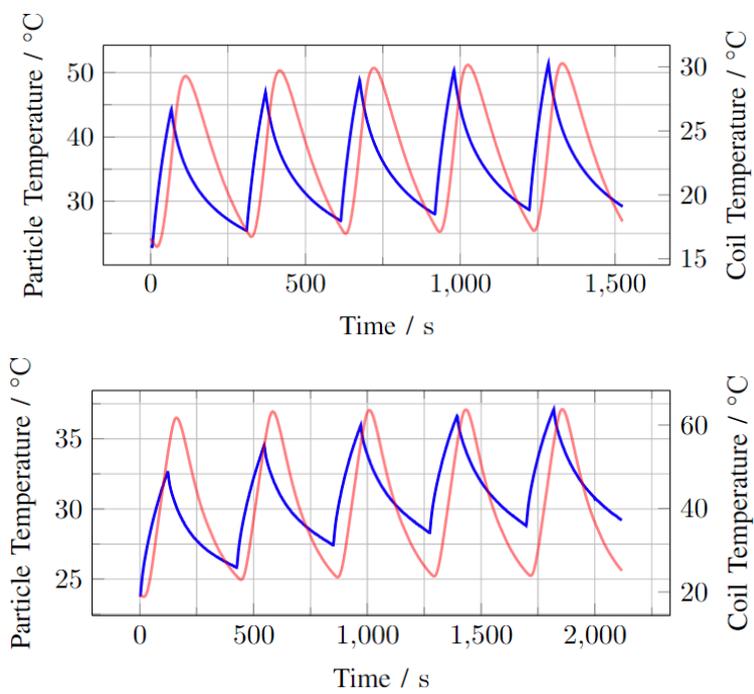

*Figure 11: Heating measurements for the single-coil MFH system (upper graph) and the split-coil MFH system (lower graph) showing 5 heating cycles and the cooling cycles demonstrating an interleaved heating scenario. Blue curve shows the particle temperature, red curve shows coil temperature.*

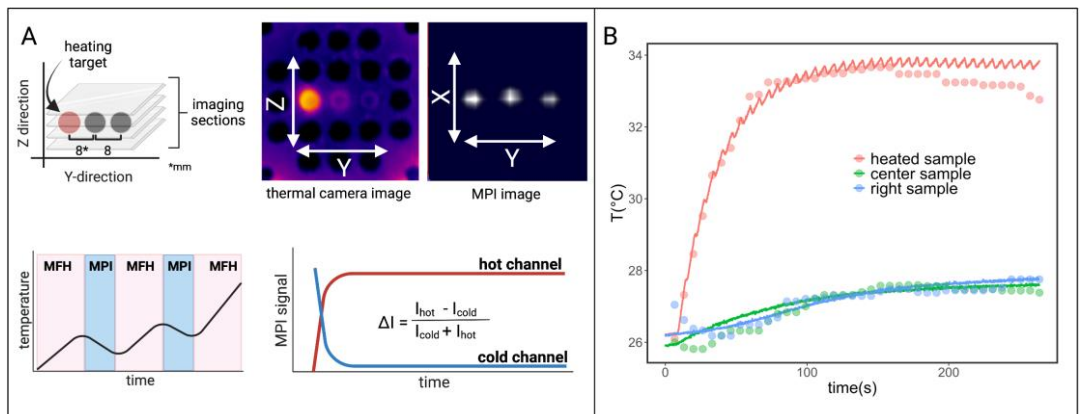

*Figure 12: Interleaved MPI and localized MFH achieved with the integrable single-coil MFH system. (A) The MPI images are reconstructed for MPI-based thermometry (B). The reconstructed temperature results (dots) are compared to the thermal camera measured temperature results (solid lines). [31]*